\documentclass[12pt]{article}

\usepackage{amssymb}

\begin{document}

\begin{titlepage}

\begin{flushright}
arXiv:1211.6614
\end{flushright}
\vskip 2.5cm

\begin{center}
{\Large \bf Lorentz and CPT Violation in\\
Scalar-Mediated Potentials}
\end{center}

\vspace{1ex}

\begin{center}
{\large Brett Altschul\footnote{{\tt baltschu@physics.sc.edu}}}

\vspace{5mm}
{\sl Department of Physics and Astronomy} \\
{\sl University of South Carolina} \\
{\sl Columbia, SC 29208} \\
\end{center}

\vspace{2.5ex}

\medskip

\centerline {\bf Abstract}

\bigskip

In Lorentz- and CPT-violating effective field theories involving scalar and spinor
fields, there exist forms of Lorentz violation that modify
only the scalar-spinor Yukawa interaction
vertices. These affect low-energy fermion and antifermion scattering processes
through modifications to the nonrelativistic Yukawa potentials. The modified
potentials involve novel combinations of momentum, spin, and Lorentz-violating
background tensors.

\bigskip

\end{titlepage}

\newpage

\section{Introduction}

In recent years, a significant amount of attention has been paid to the possibility
that the laws of physics at the most fundamental level may not respect Lorentz and
CPT symmetries exactly. While there is thus far no compelling experimental reason to
believe that Lorentz or CPT invariances are
not exact, many candidate theories of quantum
gravity suggest the possibility of such symmetry violations, at least in certain
regimes. The possibilities for symmetry breaking include
spontaneous breaking in
string theory~\cite{ref-kost18,ref-kost19} and elsewhere~\cite{ref-altschul5},
mechanisms in loop quantum gravity~\cite{ref-gambini,ref-alfaro} and
non-commutative
geometry~\cite{ref-mocioiu,ref-carroll3}, Lorentz violation through
spacetime-varying couplings~\cite{ref-kost20,ref-ferrero1}, and anomalous breaking of
Lorentz and CPT symmetries~\cite{ref-klinkhamer}.

Because any confirmed discovery of Lorentz violation would be a sure sign of
new physics---with a fundamentally different structure from anything previously
observed---this subject remains quite interesting and an active area of both
experimental and theoretical research. Most theoretical work is performed within
the context of effective quantum field theory. There is an effective field theory
known as the standard model extension (SME) that contains all possible 
translation-invariant but
Lorentz-violating operators that may be constructed out of the fields of the
standard model.
(Generalizations to include additional fields are straightforward.)
Each Lorentz-violating operator that appears in the SME Lagrangian is
parameterized by a small background tensor~\cite{ref-kost1,ref-kost2}.
If the Lorentz violation arises from spontaneous symmetry breaking, these
background tensors are essentially the vacuum expectation values of tensor-valued
fields.
Moreover, because the existence of CPT violation in a stable, unitary quantum field
theory implies that there must also be Lorentz violation~\cite{ref-greenberg},
the SME is also the most general effective field theory describing CPT violation.

The most frequently considered subset of the SME is the
minimal SME, which contains only gauge-invariant, local,
superficially renormalizable forms of Lorentz
violation. The minimal SME has become the standard framework used for parameterizing
the results of experimental Lorentz tests.
Recent searches for Lorentz violation have included studies of matter-antimatter
asymmetries for trapped charged
particles~\cite{ref-bluhm1,ref-gabirelse,ref-dehmelt1} and bound state
systems~\cite{ref-bluhm3,ref-phillips},
measurements of muon properties~\cite{ref-kost8,ref-hughes}, analyses of
the behavior of spin-polarized matter~\cite{ref-heckel3},
frequency standard comparisons~\cite{ref-berglund,ref-kost6,ref-bear,ref-wolf},
Michelson-Morley experiments with cryogenic
resonators~\cite{ref-muller3,ref-herrmann2,ref-herrmann3,ref-eisele,ref-muller2},
Doppler effect measurements~\cite{ref-saathoff,ref-lane1},
measurements of neutral meson
oscillations~\cite{ref-kost10,ref-kost7,ref-hsiung,ref-abe,
ref-link,ref-aubert}, polarization measurements on the light from cosmological
sources~\cite{ref-carroll2,ref-kost11,ref-kost21,ref-kost22},
high-energy astrophysical
tests~\cite{ref-stecker,ref-jacobson1,ref-altschul6,ref-altschul7,ref-klinkhamer2},
precision tests of gravity~\cite{ref-battat,ref-muller4}, and others.
The results of these experiments set constraints on the various SME coefficients, and
up-to-date information about
most of these constraints may be found in~\cite{ref-tables}.

The least studied areas of the SME are those that involve scalar fields.
There has been a good deal of theoretical investigation into the behavior of
scalars in the (Lorentz-invariant) standard model---both the fundamental Higgs and
composite pseudoscalar mesons in the hadronic sector.
However, little attention has been
paid to scalars by theorists studying Lorentz violation.
For example, the one-loop renormalization of the
SME Higgs sector has not yet been studied
systematically. Although the one-loop renormalization of the
Abelian~\cite{ref-kost3}, non-Abelian~\cite{ref-collad-3}, and chiral~\cite{ref-collad-1} gauge theories with
spinor matter that make up parts of the SME were completed some time ago, only
recently have the corresponding scalar field theories including Yukawa interactions~\cite{ref-ferrero3} received similar treatments. Yet the study of
the renormalization of the scalar sector is still not complete; gauge theories with
scalar matter and theories with spontaneous gauge symmetry breaking have not
been adequately examined.

This paper discusses how low-energy scalar-mediated
interactions between fermions and antifermions may be affected by Lorentz violation.
At nonrelativistic energies, these interactions are described by modified Yukawa
potentials. The changes to the Yukawa potential induced by Lorentz violation in the
pure scalar propagation sector were
previously considered in~\cite{ref-altschul31}. However, as emphasized in
\cite{ref-ferrero3}, the Yukawa sector contains Lorentz-violating operators that
modify the scalar-spinor interactions. These operators
can have a much more intricate structure than those considered
in~\cite{ref-altschul31}.

The nonrelativistic limit is most relevant in low-energy hadronic physics.
Understanding symmetry breaking in few-baryon systems is an important topic. Much of
the research in this area has focused on parity (P) violation, since this is the most
strongly broken spacetime symmetry in the standard model. Weak P violation in
processes that are normally
dominated by the strong interaction has become a very interesting
area of research, with experimental data coming from a number of different nuclear
systems~\cite{ref-holstein}. The same kinds of hadronic experiments, particularly
those involving precision measurements of neutron spin rotations, may be useful
as tests of Lorentz and CPT invariances.

P-violating observables in systems of interacting nucleons have
traditionally been analyzed in terms of interparticle potentials, as in the
Desplanques-Donoghue-Holstein (DDH) model~\cite{ref-desplanques}. The DDH model uses
a collection of one-boson exchange potentials, generated by the exchange of both
scalar and vector mesons. More recent analyses have used effective field
theories, particularly a pionless~\cite{ref-zhu} effective theory that
should be reliable at energies less than $m_{\pi}^{2}/M_{N}$.

The state of the art for P violation in field theory has moved beyond
two-nucleon systems. Three-nucleon observables, such as spin rotations in
neutron-deuteron scattering, have been analyzed using the pionless effective field
theory~\cite{ref-greisshammer1,ref-vanasse},
as well as hybrid methods that use effective field theory to describe the symmetry violation along with modified wave functions derived from the P-invariant portions of
the interparticle potentials~\cite{ref-schiavilla,ref-song1}. These field theory
methods have also been extended to deal with time reversal invariance
violation~\cite{ref-song2}, which is much weaker than P violation in the standard
model.

However, this paper will only consider two-body potentials. The rich additional
structure that arises when violations of isotropy and boost invariance are allowed
may make three-body physics more complicated (in contrast to the Lorentz-invariant
pionless theory, which has no leading-order three-body
interactions~\cite{ref-greisshammer2}). The potential theory provides a starting point
for the analysis of observables that break Lorentz invariance.
Our approach to deriving the relevant potentials will be somewhat similar to the
one used to determine the low-energy forms of P violation
in~\cite{ref-girlanda}---starting with the possible relativistic forms of symmetry
violation, then reducing to a nonrelativistic theory and observing any redundancies.

The outline of this paper is as follows. Section~\ref{sec-scalarLV} describes the
structure of the Lorentz-violating operators that appear in the scalar sector of a
Lorentz-violating effective field theory. Then section~\ref{sec-potential} examines
how Lorentz-violating modifications to scalar-spinor couplings affect the Yukawa
potentials between spin-$\frac{1}{2}$ particles. The results are summarized in
section~\ref{sec-concl}, along with some additional remarks placing this work in
context.

\section{Scalar Sector Lorentz Violation}
\label{sec-scalarLV}

The Lagrange density for the minimal SME contains local operators that can be
constructed out of the standard model's scalar, spinor, and gauge fields. To maintain
superficial renormalizability, only gauge invariant operators with mass dimensions
of up to 4 are included. In all cases that have been checked explicitly, these
conditions are indeed sufficient to make the theories renormalizable at one-loop
order. (As already noted, the Yukawa theory discussed in this paper is among the
cases for which such a check has been made.)

For a single species of Dirac fermion, the minimal SME Lagrange density is
\begin{equation}
\label{eq-Lf}
{\cal L}_{f}=\bar{\psi}(i\Gamma^{\mu}\partial_{\mu}-M)\psi,
\end{equation}
where
\begin{equation}
\Gamma^{\mu}=\gamma^{\mu}+\Gamma_{1}^{\mu}
=\gamma^{\mu}+c^{\nu\mu}\gamma_{\nu}+d^{\nu\mu}\gamma_{5}\gamma_{\nu}
+e^{\mu}+if^{\mu}\gamma_{5}+\frac{1}{2}g^{\lambda\nu\mu}
\sigma_{\lambda\nu}
\end{equation}
and
\begin{equation}
\label{eq-LM}
M=m+im_{5}\gamma_{5}+M_{1}=
m+im_{5}\gamma_{5}+a^{\mu}\gamma_{\mu}+b^{\mu}\gamma_{5}\gamma_{\mu}+
\frac{1}{2}H^{\mu\nu}\sigma_{\mu\nu}.
\end{equation}
These are the only operators satisfying the listed conditions that can exist in
a purely fermionic theory. The $\Gamma$ coefficients are dimensionless, while the
$M$ coefficients have dimension $({\rm mass})^{1}$.
However, some of the coefficients appearing in
$\Gamma$ and $M$ are more interesting than others. Several, such as $m_{5}$, $a$, and
$f$, may be eliminated from the theory by a redefinition of the fermion
field~\cite{ref-colladay2,ref-altschul8}.

The species of interest in low-energy hadronic physics are composites formed from
the fundamental quark and gluon fields. However, there are still Lorentz violation coefficients for the composite species; they are linear combinations of the
coefficients for the fundamental fields. Many of the more precise bounds on SME
coefficients are actually constraints on the composite coefficients for protons and
neutrons.

Many of the terms present in (\ref{eq-Lf}--\ref{eq-LM})
violate CPT as well as Lorentz symmetry.
Of the coefficients that cannot be eliminated by field redefinitions, those with odd
numbers of Lorentz indices are also odd under CPT. However, the coefficients that
have even numbers of indices are CPT invariant. It is thus possible to break Lorentz
symmetry but leave CPT intact (although not vice versa). The full discrete symmetry
properties of the minimal SME operators are discussed
in~\cite{ref-kost3,ref-altschul8}.

To the fermionic theory may be appended one or more boson fields---of either scalar
or gauged vector types. This also introduces new possible forms of Lorentz violation.
However, there is a fundamental difference between the possibilities in scalar- and
gauge-mediated interactions. In a gauge theory, whatever renormalizable
Lorentz violation exists in the
free fermion sector completely determines the Lorentz violation present at the
boson-fermion vertex. The same
quantity $\Gamma^{\mu}$ appears in both the fermion propagator and the vertex,
because of gauge invariance. However, the situation is quite different in a Yukawa
theory. With no additional condition analogous to gauge invariance, there
is a completely independent set of Lorentz-violating operators that can appear in
the fermion-scalar vertex.

With the addition of a scalar field $\phi$, the most general Lorentz-violating
Lagrange density that does not lead to spontaneous symmetry breaking becomes
\begin{equation}
\label{eq-Lb}
{\cal L}=\bar{\psi}(i\Gamma^{\mu}\partial_{\mu}-M)\psi+\frac{1}{2}(\partial^{\mu}
\phi)(\partial_{\mu}\phi)+\frac{1}{2}K^{\mu\nu}(\partial_{\nu}\phi)
(\partial_{\mu}\phi)-\frac{1}{2}\mu^{2}\phi^{2}-\frac{\lambda}{4!}\phi^{4}
-\bar{\psi}G\psi\phi.
\end{equation}
The symmetric tensor $K^{\mu\nu}=K^{\nu\mu}$ represents the only kind of Lorentz
violation that can be introduced in the pure bosonic sector with a real scalar field.
Much more intricate in structure is the operator $G$ appearing in the Yukawa vertex
term. $G$ has essentially the same
structure as the $M$ term in the pure scalar sector,
\begin{equation}
G=g+ig'\gamma_{5}+G_{1}=g+ig'\gamma_{5}+I^{\mu}\gamma_{\mu}+J^{\mu}\gamma_{5}
\gamma_{\mu}+\frac{1}{2}L^{\mu\nu}\sigma_{\mu\nu}.
\end{equation}
The terms $g$ and $g'$ are the usual scalar and pseudoscalar Yukawa couplings, while
the other terms are Lorentz violating. All the coefficients contained in $G$ are
dimensionless. The tensor term $L^{\mu\nu}$ is naturally
antisymmetric. The discrete symmetries of the operators that make up $G$ are similar
to the symmetries of the corresponding operators contained in $M$. If the $\phi$
field is a true scalar, the symmetries are exactly the same; for a pseudoscalar
field, the parity and time reversal behaviors are opposite between $M$ and $G$, while
the charge conjugation properties are still the same. Ultimately, the $I$ and $J$
terms violate CPT as well as Lorentz symmetry, while $L$ is CPT invariant.

While the fact that this rich structure of operators could exist in the
spinor-scalar coupling term was noted as part of the original
formulation of the SME, very little attention has been paid to the $G$ terms.
There has been essentially no calculations of their phenomenalistic effects, and only
recently~\cite{ref-ferrero3} have the effects of these terms on the
one-loop renormalization of the SME been considered.

\section{Modified Yukawa Potentials}
\label{sec-potential}

Both the $K$ and $G$ terms will affect the Yukawa potentials for interacting
fermions and antifermions, because both of them appear in the four-point correlation
functions that describe two-particle scattering. The purpose of this section will be
to evaluate the interparticle potentials that are associated with this scattering.
Since the $K$ term was already discussed in~\cite{ref-altschul31}, the focus here
will be on the effects of the $I$, $J$, and $L$ terms that, together with the
Lorentz-invariant $g$ and $g'$ terms, comprise $G$.

In discussing the modified Yukawa potential,
we shall only consider Lorentz-violating effects that are linear in the
SME coefficients. Since Lorentz violation is known to be a very small phenomenon
physically, higher-order effects should be minuscule. A similar leading-order
analysis of electromagnetic potentials was conducted in~\cite{ref-kost17}.

Lorentz violation in the pure fermion sector (and in the fermion-gauge
interaction sector) is relatively well constrained, at least for the first generation
fermions that make up the stable constituents of everyday matter. For this reason, we
shall neglect the $\Gamma_{1}$ and $M_{1}$ terms (even though some small
nonzero $\Gamma_{1}$ and $M_{1}$ terms could be generated from $G_{1}$ by radiative corrections~\cite{ref-ferrero3}).

However, if the coefficients of the
operators involved were not too small, the forms of Lorentz violation described by
$\Gamma_{1}$ and $M_{1}$ would affect fermionic scattering in a significant way.
These pure fermion sector terms
would affect both the amputated matrix element for the one-boson
exchange process that dominates low-energy scattering and the
dispersion relations for the external particles, which would in turn affect the
kinematics of a reaction. In fact, the changes to scattering and decay rates due to
changes in
particle velocities and available phase space may be as large as or larger than the
changes arising from Lorentz violation in the invariant matrix element
itself~\cite{ref-kost5,ref-altschul2,ref-altschul32}.

\subsection{Spinor Products}
\label{sec-products}

When $\Gamma_{1}$ and $M_{1}$ are neglected, it is possible to use standard external
fermion
and antifermion spinor states for the calculation of a matrix element. Since the
behaviors of the $G$ operators depend in nontrivial ways on the spins of the external
particles, it is simplest to perform the matrix element calculation using explicit
spinor eigenstates.
Using the Dirac-Pauli basis for the Dirac matrices and a relativistic normalization
convention, the Dirac spinor
$u_{s}(p)$ (corresponding to momentum $p$ and spin $s$) is
\begin{equation}
u_{s}(p)=\sqrt{\frac{2E(E+m)}{2m}}\left[
\begin{array}{c}
\xi_{s} \\
\frac{\vec{\sigma}\cdot\vec{p}}{E+m}\xi_{s}
\end{array}
\right],
\end{equation}
where, $\xi_{s}$ is a two-component spinor. In the nonrelativistic limit, this
becomes
\begin{equation}
u_{s}(p)=\sqrt{2m}\left[
\begin{array}{c}
\xi_{s} \\
\frac{\vec{\sigma}\cdot\vec{p}}{2m}\xi_{s}
\end{array}
\right].
\end{equation}
Using the explicit spinors, it is possible to calculate the fermion bilinears
that appear in a scattering amplitude. In particular, if the external particles are
nonrelativistic, so that terms with more than a single power of $p/m$ may be
neglected,
\begin{eqnarray}
\bar{u}_{s'}(p')Gu_{s}(p) & = & 2m\left[\xi^{\dag}_{s'}\,, -\xi^{\dag}_{s'}
\frac{\vec{\sigma}\cdot\vec{p}\,'}
{2m}\right]\left(g+i\gamma_{5}g'+G_{1}\right)\left[
\begin{array}{c}
\xi_{s} \\
\frac{\vec{\sigma}\cdot\vec{p}}{2m}\xi_{s}
\end{array}
\right] \\
& = & 2m(g+I_{0})\xi^{\dag}_{s'}\xi_{s}+ig'(p_{j}-p'_{j})\xi^{\dag}_{s'}
\sigma_{j}\xi_{s}+2m(J_{j}+\epsilon_{jkl}L_{kl})\xi^{\dag}_{s'}\sigma_{j}\xi_{s}
\nonumber\\
& & -I_{j}\left[(p_{j}+p'_{j})\xi^{\dag}_{s'}\xi_{s}-i\epsilon_{jkl}
(p_{k}-p'_{k})\xi^{\dag}_{s'}\sigma_{l}\xi_{s}\right]-J_{0}(p_{j}+p'_{j})
\xi^{\dag}_{s'}\sigma_{j}\xi_{s} \nonumber\\
& & -L_{0j}\left[i(p_{j}-p'_{j})\xi^{\dag}_{s'}\xi_{s}-\epsilon_{jkl}(p_{k}+p'_{k})
\xi^{\dag}_{s'}\sigma_{l}\xi_{s}\right].
\label{eq-uGu}
\end{eqnarray}
If the two fermions involved are of different species, direct scattering is the
only possible channel. We shall henceforth assume that the particles involved in a
scattering event are indeed distinguishable. However, in the scattering of identical
particles, the usual method of replacing the scattering amplitude $f(\theta,\phi)$
with $f(\theta,\phi)-f(\pi-\theta,\pi+\phi)$ will give the correct result.

For antiparticle scattering, we shall similarly assume that the annihilation
scattering channel is not available, and only a single diagram contributes to the
potential. We may also take advantage of the fact that
$v_{s}(p)=\gamma_{5}u_{-s}(p)$, where the subscript $-s$ on the spinor $u(p)$
indicates a spinor with spin projections that are opposite those of $u_{s}(p)$.
Then we have that
\begin{eqnarray}
\label{eq-vGv0}
\bar{v}_{s}(p)Gv_{s'}(p') & = & -\bar{u}_{-s}(p)\gamma_{5}G\gamma_{5}u_{-s'}(p')\\
& = & -2m(g-I_{0})\xi^{\dag}_{-s}\xi_{-s'}+ig'(p_{j}-p'_{j})\xi^{\dag}_{-s}
\sigma_{j}\xi_{-s'}+2m(J_{j}-\epsilon_{jkl}L_{kl})\xi^{\dag}_{-s}\sigma_{j}\xi_{-s'}
\nonumber\\
\label{eq-vGv}
& & -I_{j}\left[(p_{j}+p'_{j})\xi^{\dag}_{-s}\xi_{-s'}+i\epsilon_{jkl}
(p_{k}-p'_{k})\xi^{\dag}_{-s}\sigma_{l}\xi_{-s'}\right]-J_{0}(p_{j}+p'_{j})
\xi^{\dag}_{-s}\sigma_{j}\xi_{-s'} \nonumber\\
& & -L_{0j}\left[i(p_{j}-p'_{j})\xi^{\dag}_{-s}\xi_{-s'}+\epsilon_{jkl}(p_{k}+p'_{k})
\xi^{\dag}_{-s}\sigma_{l}\xi_{-s'}\right].
\end{eqnarray}
The remaining products of Pauli spinors can be simplified further. The inner product
$\xi^{\dag}_{-s}\xi_{-s'}$ simply equals $\xi^{\dag}_{s'}\xi_{s}=\delta_{ss'}$.
However, because of the spin reversal present in $\xi_{-s}$, the matrix element
$\xi^{\dag}_{-s}\sigma_{j}\xi_{-s'}$ is equal to $-\xi^{\dag}_{s'}\sigma_{j}\xi_{s}$.

Note that present in (\ref{eq-uGu}) [and (\ref{eq-vGv})]
are almost all the possible vector structures that
can be constructed at first order in the momenta. There are three independent
three-vector
operators that may be constructed:  the total momenta along the incoming and
outgoing lines from a vertex ($\vec{p}+\vec{p}\,'$), the momentum transfer
($\vec{q}=\vec{p}-\vec{p}\,'$), and the spin operator ($\vec{\sigma}$). Each of these
three may form a dot product with an isotropy-breaking background vector; either of
the momentum observables may form a dot product with the spin; or there may be a
triple product with a background vector, the spin, and one of the momenta. The
only possible structures that are missing are contractions of the momenta and spin
with symmetric, traceless three-tensors;
however, these structures cannot exist because there is no
such symmetric, traceless tensor that can be constructed at first order in $G$.

The terms involving background three-vectors manifestly break isotropy. Moreover,
Lor\-entz boost invariance normally prevents the appearance of the
$\vec{p}+\vec{p}\,'$ terms. The average velocity 
$\vec{v}_{av}=\frac{1}{2}(\vec{v}+\vec{v}\,')$ (which is, of course, the passing
velocity $\vec{v}_{av}\approx\vec{v}\approx\vec{v}\,'$ in a glancing collision)
is measured in the specific laboratory frame in which the calculation has been
performed. The dot product of this velocity with the spin is not invariant under
nonrelativistic Galilean transformations, and this signifies the failure of Lorentz
boost symmetry. In contrast, the difference $\vec{v}-\vec{v}\,'$, which
appears in the Lorentz-invariant theory in conjunction with the pseudoscalar $g'$
term, transforms covariantly under Galielan transformations


\subsection{Fermion Potential}

For the scattering of two nonidentical fermions, with the exchange of a single scalar
boson between them, the matrix element is
\begin{equation}
\label{eq-M}
i{\cal M}=\bar{u}^{a}_{s_{a}'}(p'_{a})G^{a}u^{a}_{s_{a}}(p_{a})
\frac{-i}{q^{2}-\mu^{2}+i\epsilon}
\bar{u}^{b}_{s_{b}'}(p'_{b})G^{b}u^{b}_{s_{b}}(p_{b}).
\end{equation}
The indices $a$ and $b$ denote the identities of the species involved. However,
for simplicity, we shall assume that there is only Lorentz violation for one
species of particle (the one labeled $a$). Including Lorentz violation for both is a
straightforward generalization. We shall consider both scalar and pseudoscalar
couplings for the second particle, however.

For
nonrelativistic scattering, the interaction may be described using a potential, such
that
\begin{equation}
V(\vec{r}\,)=\int\frac{d^{3}q}{(2\pi)^{3}}e^{i\vec{q}\cdot\vec{r}}i
{\cal M}(\vec{q}\,),
\end{equation}
in the limit where $q_{0}=0$. The integrand involves the usual scalar Yukawa
amplitude proportional to $1/(\vec{q}\,^{2}+\mu^{2})$, as well as terms with
additional factors of $q_{j}$.
The spatial extent of the interactions is therefore determined by the Yukawa potential
function and its derivative,
\begin{eqnarray}
f(\vec{r}\,) & = & -\frac{e^{-\mu r}}{4\pi r} \\
g_{j}(\vec{r}\,) & = & \partial_{j}f(\vec{r}\,)=\frac{e^{-\mu r}}{4\pi r^{2}}
\left(\mu+\frac{1}{r}\right)x_{j}.
\end{eqnarray}

Although higher powers of $\vec{q}$ (and thus additional spatial derivatives) will
be largely neglected in this paper, we notice that the next order term is
\begin{equation}
h_{jk}(\vec{r}\,)=\partial_{j}\partial_{k}f(r)=\frac{e^{-\mu r}}{4\pi r^{2}}
\left[\left(\mu+\frac{1}{r}\right)\delta_{jk}-\left(\frac{\mu^{2}}{r}+
\frac{3\mu}{r^{2}}+\frac{3}{r^{3}}\right)x_{j}x_{k}\right]+\frac{1}{3}\delta^{3}
(\vec{r}\,)\delta_{jk}.
\end{equation}
Spatial potentials with this shape already appear
in the Lorentz-invariant theory, coming from terms
with pseudoscalar $g'$ couplings at both vertices. It is also possible to have a
first-order
Lorentz-violating potential with this shape, if one vertex involves a $G_{1}$ term
and the other vertex a $g'$.

The function $f$ contributes to that part of the potential that comes from vertex
terms in which $q$ does not appear. These generate a potential
\begin{equation}
\label{eq-Vf}
V_{f}(\vec{r}\,)=[\tilde{g}-I_{j}(v^{a}_{av})_{j}-J_{0}(v^{a}_{av})_{j}
\sigma^{a}_{j}+\tilde{J}_{j}\sigma^{a}_{j}
+\epsilon_{jkl}\tilde{L}_{j}(v_{av}^{a})_{k}\sigma^{a}_{l}]g_{b}f(\vec{r}\,).
\end{equation}
The terms in square in brackets in (\ref{eq-Vf}) comes from the vertex with the
$a$ species, while the $g_{b}$ term comes from the $b$ vertex.
The species labels are omitted on the
Lorentz violation coefficients, since there is assumed to be no Lorentz violation
in the vertex with the $b$ species. The scalar $\tilde{g}$ and three-vectors
$\tilde{J}_{j}$ and $\tilde{L}_{j}$ marked with tildes denote the
combinations
\begin{eqnarray}
\label{eq-gtilde}
\tilde{g} & = & g+I_{0} \\
\label{eq-Jtilde}
\tilde{J}_{j} & = & J_{j}+\epsilon_{jkl}L_{kl} \\
\label{eq-Ltilde}
\tilde{L}_{j} & = & L_{0j}=-L_{j0},
\end{eqnarray}
which are the combinations that are observable in experiments with nonrelativistic
fer\-mi\-ons
conducted in a single Lorentz frame. They are analogous to the tilde-marked
coefficients defined in other sectors of the SME, although the ones defined
in~(\ref{eq-gtilde}--\ref{eq-Ltilde}) are defined to be
dimensionless, unlike the tilde coefficients in other sectors, which most typically
have units of mass. These combinations of indistinguishable terms
exist because the leading order
fermionic matrix elements of even Dirac operators
(which involve only the large components of the Dirac spinors) are unchanged when
multiplied by $\gamma_{0}$. So if ${\cal E}$ is an even operator, the
nonrelativistic matrix elements of ${\cal E}$ and $\gamma_{0}{\cal E}$ are identical.
Note however, that different combinations exist for antifermions, because
$\gamma_{0}$ is equivalent to $-1$ in the corresponding matrix elements for
antiparticles.

Despite the existence of these degenerate combinations, the structure of the physical
matrix elements of $G$ is still quite a bit
richer than the corresponding structure for matrix elements
of $M$. Because $M$ appears in the bilinear propagation Lagrangian for fermion
species, its physical matrix elements always involved particles with identical
incoming and outgoing momenta. So a matrix element such as
$\bar{u}_{s'}(p)Mu_{s}(p)$ lacks any terms that depend on $\vec{p}-\vec{p}\,'$.

The part of the interaction potential with the more complicated $g_{j}(\vec{r}\,)$
spatial dependence arises from those terms in the scattering amplitude with just
a single factor of
$q_{j}$. This factor can appear at either vertex. At the $a$ vertex, either a
Lorentz-violating term or the $g'_{a}$ coupling may be responsible; or at the $b$
vertex, there may be a coupling $g'_{b}$. Taken together, these terms generate a
potential
\begin{eqnarray}
V_{g}(\vec{r}\,) & = & \left\{
\frac{1}{2m_{a}}\left[g'_{a}\sigma^{a}_{j}-\tilde{L}_{j}-\epsilon_{jkl}
I_{k}\sigma^{a}_{l}\right]g_{b}-\frac{1}{2m_{b}}
\left[\tilde{g}-I_{k}(v^{a}_{av})_{k}\frac{}{}\right.\right. \nonumber\\
& & \left.\left.-J_{0}(v^{a}_{av})_{k}
\sigma^{a}_{k}+\tilde{J}_{k}\sigma^{a}_{k}
+\epsilon_{kln}\tilde{L}_{k}(v_{av}^{a})_{l}\sigma^{a}_{n}\right]g'_{b}\sigma^{b}_{j}
\right\}g_{j}(\vec{r}\,).
\end{eqnarray}

For completeness, we may mention the potential term that arises when a factor of $q$
appears at both vertices. This potential has the spatial shape $h_{jk}(\vec{r}\,)$,
so that
\begin{equation}
V_{h}(\vec{r}\,)=\frac{1}{4m_{a}m_{b}}\left[g'_{a}\sigma^{a}_{j}-\tilde{L}_{j}
-\epsilon_{jln}I_{l}\sigma^{a}_{n}\right]\left(g'_{b}\sigma_{k}^{b}\right)h_{jk}
(\vec{r}\,).
\end{equation}
However, this is not a complete description of the potential at this order. Other
terms with multiple factors of $p/m$ have also been neglected (for example, in the
normalization of the Dirac spinors).

\subsection{Antifermion Potential}

The one-boson scalar exchange also generates potentials between antifermions and
other particles. We shall now look explicitly at the case where the species-$a$
particle is an antifermion, while the species-$b$ particle remains a fermion. The two
species are still different, so there is no annihilation scattering. In this
case, the potential is derived from a matrix element similar to (\ref{eq-M}):
\begin{equation}
i{\cal M}=-\bar{v}^{a}_{s_{a}}(p_{a})G^{a}v^{a}_{s_{a}'}(p_{a}')
\frac{-i}{q^{2}-\mu^{2}+i\epsilon}
\bar{u}^{b}_{s_{b}'}(p'_{b})G^{b}u^{b}_{s_{b}}(p_{b}),
\end{equation}
with the usual overall minus sign for antiparticle scattering, coming from
the fields' anticommutation. This cancels the overall minus sign from (\ref{eq-vGv0}).

There are additional minus signs associated with the $I$ and $L$ terms. These can be
read off directly from the explicit expression $(\ref{eq-vGv})$. However, they can be
derived most straightforwardly simply by noting that exactly those
operators associated with $I$ and $L$ are odd under fermion-antifermion charge
conjugation.

Therefore, the potentials, up to linear order in $p/m$, are
\begin{eqnarray}
V_{f}^{*}(\vec{r}\,) & = & [\tilde{g}^{*}+I_{j}(v^{a}_{av})_{j}-J_{0}(v^{a}_{av})_{j}
\sigma^{a}_{j}+\tilde{J}^{*}_{j}\sigma^{a}_{j}
-\epsilon_{jkl}\tilde{L}_{j}(v_{av}^{a})_{k}\sigma^{a}_{l}]g_{b}f(\vec{r}\,) \\
V_{g}^{*}(\vec{r}\,) & = & \left\{
\frac{1}{2m_{a}}\left[g'_{a}\sigma^{a}_{j}+\tilde{L}_{j}+\epsilon_{jkl}
I_{k}\sigma^{a}_{l}\right]g_{b}-\frac{1}{2m_{b}}
\left[\tilde{g}^{*}+I_{k}(v^{a}_{av})_{k}\frac{}{}\right.\right. \nonumber\\
& & \left.\left.-J_{0}(v^{a}_{av})_{k}
\sigma^{a}_{k}+\tilde{J}^{*}_{k}\sigma^{a}_{k}
-\epsilon_{kln}\tilde{L}_{k}(v_{av}^{a})_{l}\sigma^{a}_{n}\right]g'_{b}\sigma^{b}_{j}
\right\}g_{j}(\vec{r}\,).
\end{eqnarray}
Because of the sign changes associated with charge conjugation, two additional
linear combinations of coefficients are relevant for nonrelativistic experiments
with antiparticles:
\begin{eqnarray}
\tilde{g}^{*} & = & g-I_{0} \\
\tilde{J}_{j}^{*} & = & J_{j}-\epsilon_{jkl}L_{kl},
\end{eqnarray}
where the star superscript notation continues to follow~\cite{ref-tables}.

\section{Conclusions}
\label{sec-concl}

The potential $V_{f}+V_{g}$ provides a description of the ${\cal O}(p/m)$
nonrelativistic interactions between two fermion species---one with Lorentz violation
present in the scalar-spinor vertex and one without. For a Lorentz-invariant
fermion and Lorentz-violating antifermion, the equivalent potential is
$V_{f}^{*}+V_{g}^{*}$, which is related by charge conjugation in the Lorentz-violating
$a$ sector. The Lorentz-violating structure of these terms is evident in several ways.
Dependences on specific projections of the spin and momenta break spatial isotropy,
and structures such as $\vec{v}_{av}\cdot\vec{\sigma}$ break boost invariance.

This study has already demonstrated several important properties
of the Lorentz-violating operators that form $G_{1}$. In most sectors of the SME,
there are operators that are not physically observable. However, all the terms
that compose $G$ appear in the nonrelativistic potentials, making observable
contributions to the energy. While it may not be surprising that terms from $G$,
which only affect interactions, not free particle propagation, cannot be eliminated
from observables in the same fashion as $m_{5}$, $a$, or $f$, neither is it obvious
that such is the case.

The potentials derived in this paper provide a fairly general formalism for
studying violations of fundamental symmetries in low-energy, potential-dominated
interaction processes. As noted in section~\ref{sec-products}, most of the observables
than can be constructed at ${\cal O}(p/m)$ are included in the potentials, and those
that are not included cannot descend from a renormalizable relativistic quantum field
theory. The study of symmetry violation in low-energy processes is an active area of
hadronic research, and it may be
possible to place constraints on completely new SME parameters through studies of
meson-mediated interactions between baryons.

Naturally, further generalizations of the results discussed here are also possible.
There may be Lorentz violation at both vertices, and
accounting for this possibility is entirely straightforward, as are accounting for
the additional diagrams that appear when the external particles are associated with
the same species. However, it is possible that the $V_{f}$ and $V_{g}$ potentials may
not actually include the
predominant effects, even when the momentum transfer in a collision is very low.
The $g'_{a}g'_{b}$ term in $V_{h}$ is ${\cal O}(p^{2}/m^{2})$, but it is Lorentz invariant; so it would be no surprise if that term were substantially larger than
Lorentz-violating terms that are nominally lower order in $p/m$. The $g'_{a}g'_{b}$
term is in fact the dominant term in standard model interactions
involving pseudoscalar mesons when P violation is small.

Lorentz violation for the external fermion states has also
been neglected, although if such
Lorentz violation exists, it will modify the interparticle potentials further.
The purely fermionic $\Gamma_{1}$ and $M_{1}$
terms in the SME Lagrangian were neglected
because such terms, which would affect freely propagating fermions, are rather well
constrained for first-generation species. However, Lorentz violation in the scalar
sector is a separate matter, and the effects of Lorentz violation in the scalar
sector on the Yukawa potential have already been studied~\cite{ref-altschul31}. The
effect of a (CPT-even) tensor $K_{\mu\nu}$ is to modify $f(r)$ to
\begin{equation}
f^{K}(\vec{r}\,)=-\frac{e^{-\mu r}}{4\pi r}\left[1+\frac{1}{2}K_{jj}
-\frac{1}{2}K_{jk}\left(\frac{\mu}{r}+\frac{1}{r^{2}}\right)x_{j}x_{k}\right].
\end{equation}
When the vertex interactions involve the Lorentz-invariant $g'$ term, $K$
leads to a further modified version of the derivative
\begin{eqnarray}
g^{K}_{j}(\vec{r}\,) & = & \partial_{j}f^{K}(\vec{r}\,)=g_{j}(\vec{r}\,)
\left[1+\frac{1}{2}K_{kk}
-\frac{1}{2}K_{kl}\left(\frac{\mu}{r}+\frac{1}{r^{2}}\right)x_{k}x_{l}\right]
\nonumber\\
& & +\frac{1}{2}f(\vec{r}\,)K_{kl}\left[\left(\frac{\mu}{r^{3}}+\frac{2}{r^{4}}\right)
x_{j}x_{k}x_{l}-\left(\frac{\mu}{r}+\frac{1}{r^{2}}\right)(\delta_{jk}x_{l}+
\delta_{jl}x_{k})\right].
\end{eqnarray}
The modified $g_{j}^{K}(\vec{r}\,)$ is not needed in the $\vec{q}$-dependent terms
that are themselves Lorentz violating; any resulting changes to the potentials
calculated in section~\ref{sec-potential} would be higher order in the small
Lorentz violation coefficients.

%

Finally, this work provides another stepping stone in the general analysis of
Lorentz violation in scalar theories.
With the advent of the Large Hadron Collider, it appears
that it is finally possible to see direct evidence of the Higgs
boson~\cite{ref-ATLAS1,ref-CMS1}.
This naturally opens up the possibility of studying the
Lorentz symmetry behavior of fundamental scalar fields. The era of
direct experimental studies of the Higgs particle is just beginning, and the
theoretical foundation for understanding such studies needs to be prepared.
While spontaneous gauge symmetry breaking is one of the most important features of the
standard model, its complexity has limited studies of SME scalar fields to particular
sub-topics. In addition to the renormalization of the Yukawa and pure scalar sectors,
the tree-level quantization of the theory in the spontaneously broken
phase~\cite{ref-anderson,ref-altschul30} has already been studied. Certain specific
quantum corrections, originating in the Faddeev-Popov ghost
sector~\cite{ref-altschul25}, or involving higher powers of the SME
coefficients~\cite{ref-gomes2}, have also been examined. This paper presents some
new results that increase our understanding of the scalar sector of the SME. While
particles interacting via direct Higgs exchange are generally not at low
nonrelativistic energies, this work nonetheless gives a new window into the dynamics
of Higgs interactions.

\end{document}